# Revised version of Manuscript submitted to Acta Biomaterialia on October 2017

# Multifunctional pH sensitive 3D scaffolds for treatment and prevention of bone infection

Mónica Cicuéndez[1,2], Juan C. Doadrio[1], Ana Hernández[1], M. Teresa Portolés[3], Isabel Izquierdo-Barba[*,1,2], María Vallet-Regí,[*,1,2]

[1]Departamento de Química Inorgánica y Bioinorgánica, Facultad de Farmacia, Universidad Complutense de Madrid, Instituto de Investigación Sanitaria Hospital 12 de Octubre i+12, Plaza Ramón y Cajal s/n, 28040 Madrid, Spain.

[2]CIBER de Bioingeniería, Biomateriales y Nanomedicina, CIBER-BBN, Madrid, Spain.

[3]Departamento de Bioquímica and Biología Molecular I, Facultad de Ciencias Químicas, Universidad Complutense de Madrid, Instituto de Investigación Sanitaria San Carlos IdISSC, Ciudad Universitaria s/n, 28040-Madrid, Spain.

* Corresponding authors. E-mail address: ibarba@ucm.es; vallet@ucm.es;

Phone: +34913941861    Fax: +34 394 17 86

**ABSTRACT**

Multifunctional-therapeutic 3D scaffolds have been prepared. These biomaterials are able to destroy the *S. aureus* bacteria biofilm and to allow bone regeneration at the same time. The present study is focused on the design of pH sensitive 3D hierarchical meso-macroporous scaffolds based on MGHA nanocomposite formed by a mesostructured glassy network with embedded hydroxyapatite nanoparticles, whose mesopores have been loaded with levofloxacin as antibacterial agent. These 3D platforms exhibit controlled and pH-dependent levofloxacin release, sustained over time at physiological pH (7.4) and notably increased at infection pH (6.7 and 5.5), which is due to the different interaction rate between diverse levofloxacin species and the silica matrix. These 3D systems are able to inhibit the *S. aureus* growth and to destroy the bacterial biofilm without cytotoxic effects on human osteoblasts and allowing an adequate colonization and differentiation of preosteoblastic cells on their surface. These findings suggest promising applications of these hierarchical MGHA nanocomposite scaffolds for the treatment and prevention of bone infection.

# 1. INTRODUCTION

Osteomyelitis (OM) is a bone infection with very important clinical and socio-economic implications, which is mainly caused by the pathogen *Staphylococcus aureus* (*S. aureus*). OM results in inflammatory reaction that leads to bone destruction (osteolysis), being extremely difficult to treat and may lead to patient death. [1-4]. Bacteria reach the bone (usually medular one) through hematogenous route or by direct inoculation (trauma) developing an acute infection with polymorphonuclear infiltrates, cytokines and an inflammatory reaction. Then, a cavity (*sequestrum*) is provoked, where they form a biofilm inside the bone. This biofilm involves a bacterial community embedded within a dense and protective extracellular matrix, which constitutes a primary barrier in its treating. Thus, bacteria within a biofilm can evade both the host immunological response and the action of antimicrobial agents [5-7]. OM treatment remains a significant clinical challenge due to the disadvantages presented by current therapies. Such therapies involve systemic antibiotics administration and surgery, with serious repercussions for the patients as high incidence of side effects, prolonged hospital stays, and even, high morbidity rate [8-11]. Recently, the scientific efforts are directed to the design of 3D scaffolding that dynamically contribute to the regeneration process interfering with the host body response, promoting integration, osteoconduction and angiogenesis, avoiding bacterial infection and/or offering many other desired functions to promote faster and secure healing [12-17]. Despite all the advancements, there are still challenges to be addressed in the field of local bone drug delivery including effective and sustained release control, prolonged drug stability and activity as well as the cell toxicity [18,19]. Idealistically, the local release from 3D scaffolds should preserve the stability of the loaded active molecules over time and ensure precise control over the drugs release rate [20]. That can be interpreted as going towards smart delivery of the drugs. In these cases, pH responsive approaches can allow a selective release approach either through changing the solubility of the drug carrier or cleavage of pH responsive bonds upon variation of microenvironment pH [21,22]. Such pH variations, from physiological pH (7.4) to pH values between 6.7 and 5.5, are due to the production of metabolites as lactic acid by the microorganism proliferation [23,24]. However, these pH responsive devices, in the most of the cases do not response to these variations, detecting more abrupt pH changes [25].

Recently, hierarchical meso-macroporous 3D scaffolds based on nanocomposite formed by nanocrystalline apatite uniformly embedded into a mesostructured $SiO_2$–$CaO$–$P_2O_5$ glass wall (MGHA) [26-28] have been fabricated by rapid prototyping technique. These 3D scaffolds have shown excellent properties as highly bioactive behavior, enhanced biocompatibility, no induction of inflammatory response, as well as good preosteoblast adhesion, colonization, proliferation and differentiation (see supporting information Fig.S1). Unlike other reported 3D systems [12-17,29,30], these MGHA scaffolds are made up exclusively of pure ceramic nanostructured material and not ceramic-polymer combinations which allow for excellent bioactivity together with greater homogeneity of these platforms. All these characteristics suggest their great potential application as multifunctional biomaterials for both bone tissue engineering and local drug delivery systems. Herein, with the aim of using these MGHA 3D scaffolds for treatment and prevention of OM, these 3D hierarchical platforms have been loaded with levofloxacin (Levo). This quinolone antibiotic has been chosen because: (i) it is a broad spectrum antibiotic used in bone infection therapy, which penetrates into both trabecular and cortical bone, (ii) it minimizes the risk of bacteria resistance and (iii) it exhibits different protonation states (cationic, *zwitterionic*, and anionic states) as a function of pH, which could lead to different interactions between Levo and silanol groups of the mesoporous matrix [31-33]. Thus, the present study is focused on the design of multifunctional MGHA 3D scaffolds with the ability of pH-releasing Levo to inhibit the *S. aureus* growth and to destroy the biofilm, and at the same time allowing an adequate osteoblast colonization. For these purposes, in vitro Levo release has been performed at different pHs to confirm the pH-dependent release. The effectiveness of these 3D hierarchical systems has been evaluated by both in vitro bacterium/bone cell co-culture model and on preformed Gram-positive biofilms. Finally, the possible cytotoxic effects induced by the highest Levo dose released from MGHA 3D scaffolds were evaluated with cultured human Saos-2 osteoblasts. In the present study, MGHA 3D-Levo scaffolds have been designed with two objectives: to prevent bacterial contamination and to treat bone defects caused by OM infection.

## 2. MATERIALS AND METHODS

### 2.1 Preparation and characterization of MGHA scaffolds

MGHA scaffolds based on nanocomposite material were prepared by rapid prototyping (RP) using methylcellulose (MC), as previously reported [27]. These 3D scaffolds have been characterized by X-Ray diffraction (XRD) in a Philips X´Pert diffractometer (Eindhoven, The Netherlands), equipped with Cu Kα (40 kV, 20 mA). The textural properties were determined by $N_2$ adsorption porosimetry using a Micromeritics ASAP2020 analyzer (Norcross, USA). The surface area ($S_{BET}$) was determined using the multipoint Brunauer-Emmett-Teller method included in the software. The total pore volume (Vp) was calculated from the amount of $N_2$ adsorbed at a relative pressure of 0.97. The average mesopore diameter (Dp) was obtained from the adsorption branch of the isotherm by means of the Barrett–Joyner–Halenda (BJH) method [34] Morphological structure analysis was performed by scanning electron microscopy (SEM) using a field emission JEOL JSM 6335F microscope (Tokyo, Japan) at an acceleration voltage of 10 kV. The nanostructure was characterized by Transmission electron microscopy (TEM) was performed on a JEOL 3010 electron microscope (Jeol Ltd., Japan) operating at 300 kV (Cs; 0.6 mm, resolution 1.7 °A). The porosity was measured by Hg intrusion porosimetry in a Micromeritics Autopore IV 9500 device (Micromeritics Instrument Corporation, Norcross, GA, USA). Elemental analyses (C, H, N) were carried out on a LECO CHNS-932 microanalyzer (Saint Joseph, Michigan, USA). Fourier Transform Infrared (FTIR) spectroscopy was performed in a Thermo Nicolet Nexus spectrometer (Thermo Scientific, USA) from 4000 to 400 $cm^{-1}$, using the KBr pellet method and operating in transmittance mode.

### 2.2 Levofloxacin loading into MGHA scaffolds

Levo incorporation into the 3D scaffold mesoporous structure was carried out by the impregnation method, soaking individually each scaffold (with a mass of 23 ± 2 mg) in 3 mL of an ethanol-Levo solution [37 mg/mL], at room temperature, as previous results in powder [35]. After 24 h in orbital shaking at 200 r.p.m, the scaffolds were removed, vigorous washed with ethanol and dried at room temperature. The handling of these scaffolds was performed in the dark conditions due to its sensitivity to light. The amount of drug incorporated was determined by elemental chemical analysis. Each load measurement was performed in quadruplicate. The scaffolds loaded with levofloxacin have been named as MGHA-Levo scaffolds.

**2.3 Levofloxacin release from MGHA-Levo scaffolds**

The kinetic studies of Levo release have been carried out in phosphate buffered saline (PBS) at different pHs: 5.5, 6.7 and 7.4, to study the pH influence on the release kinetics. In these studies, each MGHA-Levo scaffold (supported vertically with a Pt wire) has been soaked in 15 mL of PBS, with the pH adjusted previously, at 37ºC and orbital shaking at 300 r.p.m. The media have been removed daily. The Levo released has been quantified using a spectrofluorimeterBiotekPowerwave XS, version 1.00.14 of the Gen5 program, with a λexcitation = 292 nm and λemmision = 494 nm (26). Different calibration lines have been calculated at different pHs in a concentration range of 12 to 0.01 µg/mL.

**2.4 Antimicrobial activity of the MGHA-Levo scaffolds**

Prior to the all *in vitro* tests, the samples were sterilized by UV light radiation during 10 min in both sides. Then the samples were immersed in 6 mL of PBS under a 5% $CO_2$ atmosphere at 37ºC for 1 h to become stabilized and eliminate some residual agents before the *in vitro* culture. The efficacy against bacterial growth and biofilm has been determined using the *S. aureus*15981 laboratory strain due to the most of the bone infections are caused by this kind of bacteria [16]. Briefly, bacteria were inoculated in tryptic soy broth (TSB; BioMerieux, Marcy L'Etoile, France) and incubated overnight at 37ºC with orbital shaking at 200 r.p.m. After culture, bacteria were centrifuged for 10 min at 3500 g at 22ºC. The supernatant was then discarded and the pellet was washed three times with sterile PBS. The bacteria were then suspended and diluted in PBS to obtain a concentration of $10^8$ colony-forming units (CFU) per mL; the bacterial concentration was determined by spectrophotometry using a visible spectrophotometer (Photoanalizer D-105, Dinko instruments). To determine their effectiveness over time against *S. aureus* growth, the 3D scaffolds were soaked in 2 mL of PBS containing bacteria at $10^8$ CFU/mL of concentration. This solution was removed daily during a period of time of 1-15 days. The presence or not of bacteria, as well as their quantification, was determined on the PBS-solution in contact with the 3D-scaffolds. The determination of bacteria amount was performed by CFU in agar. In this sense, 10 µL of this solution was seeded onto TH agar and incubated at 37ºC overnight and posterior count. The minimum inhibitory concentration (MIC) of Levo has been also calculated, obtaining a value of 0.06 µg/mL. Moreover, a direct observation of 3D scaffold surfaces after different times of incubation was also performed by confocal microscope in a Biorad MC1025 microscope. For these experiments, the samples were stained for 15 min with a Live/Dead® Bacterial Viability Kit (BacklightTM). Staining was performed with a mixture of dyes: SYTO 9 (live bacteria/green) and propidium iodide (dead bacteria/red). SYTO 9 fluorescence was excited at 480/500 nm, and the emitted fluorescence measured at 500 nm, and propidium iodide fluorescence was excited at 490/635 nm, with the emitted fluorescence measured at 618 nm [36]. Controls with MGHA scaffolds without levofloxacin were always carried out in all the experiments.

Moreover, the effectiveness of these systems against biofilm, previously formed onto cover glass disks, was also determined. For these purposes, *S. aureus* biofilms were developed by suspended cover glass disks in a bacteria solution of $10^8$ bacteria per mL during 48 h at 37ºC and orbital stirring at 100 r.p.m. In this case, the medium used was 66% TSB + 0.2% glucose to promote robust biofilm formation. After that, the cover glass disks containing biofilm were localized onto six well culture plates (CULTEK) in 6 mL of new medium. Then, 3D scaffolds were submerged avoiding the direct contact with biofilm coated glass disk. After 24 h of incubation, the glass-disk were washed three times with sterile PBS, stained with a 3 μL/mL of Live/Dead® Bacterial Viability Kit (BacklightTM) and 5 μL/mL of calcofluor solution to specifically determine the biofilm formation, staining the mucopolysaccharides of the biofilm (extracellular matrix in blue). Both reactants were incubated for 15 min at room temperature. Biofilm formation was examined in an Olympus FV1200 confocal microscope [36].

**2.5 *In vitro* co-culture studies. MC3T3-E1 preosteoblasts *vs Staphylococcus aureus***

Since multifunctional 3D systems involve both antibiotic release and bone regeneration processes, the need for a flexible in vitro test system highly representative of in vivo conditions is of paramount importance before carrying out any tests on animals. Such model system allows more realistic assessment of different clinical treatment options in a rapid, cost-efficient, and safe manner, especially regarding to test possibly host-toxic therapies [37,38]. Here, we have established an in vitro bacterium/bone cell co-culture model system to evaluate these 3D systems for the prevention and treatment of OM. A preliminary, co-culture assays was carried out by seeding MC3T3-E1 cells (a murine calvaria-derived pre-osteoblastic cell line used as an archetypal model of *in vitro* osteogenesis [39, 40]) and *S. aureus* 15981 laboratory strain on the 3D-scaffolds (MGHA and MGHA-Levo), concurrently performing the corresponding control studies of osteoblasts and bacteria individually. First of all, the preosteoblastic cells were grown in complete Dulbecco's Modified Eagle's Medium (α-DMEM, Sigma Chemical Company, St. Louis, MO, USA) supplemented with 2 mM L-glutamine (BioWhittaker), 100 µg·ml$^{-1}$ penicillin (BioWhittaker), 100 g·ml$^{-1}$ streptomycin (BioWhittaker) and fetal bovine serum (FBS, Gibco) at 10% at 37 °C under atmosphere conditions of 95% humidity and 5% $CO_2$. Medium was changed every day until confluence reached ≈ 90%. Cells were washed with PBS, harvested using 0.25% trypsin-EDTA solution (Sigma Aldrich), counted with a Neubauer Hemocytometer and collected. Cells were then centrifuged at 310 g for 10 min and suspended in fresh medium without antibiotics at a density of $2 \cdot 10^4$ cells/mL. Simultaneously, *S. aureus* bacteria were cultured in the same conditions and taken at a density of $2 \times 10^8$ cells/mL, mimicking an infectious process. Then 500 µL of both preosteoblast and bacteria suspensions were seeded onto 3D scaffolds and incubated different times at 37 °C under atmosphere conditions of 95% humidity and 5% $CO_2$. Then, the samples were washed twice with sterile PBS and fixed with paraformaldehyde (Sigma Aldrich) 4% + 1% (w/v) sucrose (Sigma Aldrich) in PBS for 40 minutes. Afterwards, samples were again washed with PBS and permeabilized with 0.5% Triton X-100 (Sigma Aldrich) at 4 °C for 5 min. Nonspecific union sites were blocked with 1% (w/v) bovine serum albumin (BSA) (Sigma Aldrich) incubated at 37 °C for 20 min. Finally, samples were stained with Atto 565 conjugated with phalloidin (ratio 1:40, Sigma Aldrich) for cytoplasm α-F-actin filaments dying and then stained 2 min with 100 µl of DAPI (Thermo Fisher Scientific) for eukaryotic nuclei visualization and bacteria. Finally, the samples were washed with PBS and maintained in PBS for confocal analysis. Moreover, lactate dehydrogenase activity (LDH) was determined in the culture medium in contact with the 3D scaffolds after 6, 24 h and 3 days of incubation to evaluate the plasma membrane integrity. Activity of LDH released by the MC3T3-E1 cells is directly related to the rupture of the plasma membrane that produces the release of enzymes present in the cytoplasm. Measurements were performed by using a commercial kit (Spinreact) at 340 nm with a Beckman DU 640 UV-Visible spectrophotometer Unicam UV-500 and calculated in terms of absorbance variation by minute to obtain the units per litre concentration (ΔA/min 4925 = U / L LDH).

**2.6 Culture cell for the *in vitro* biocompatibility studies**

The *in vitro* biocompatibility assays were carried out on human osteosarcoma Saos-2 cell line which is usually used as experimental model in this kind of in vitro studies due to its osteoblastic properties as production of mineralized matrix, high alkaline phosphatase levels, PTH receptors and osteonectin presence [41]. Human Saos-2 osteoblasts (pursached from American Type Culture Collection, ATCC) were seeded on six well culture plates (CULTEK), at a density of $10^5$ cell/mL in DMEM supplemented with 10% fetal bovine serum (FBS, Gibco), 1mM L-glutamine (BioWhittaker), penicillin (200 μg/mL, BioWhittaker), and streptomycin (200 μg/mL, BioWhittaker), under a $CO_2$ (5%) atmosphere and at 37ºC for 24 h. Saos-2 is a human osteosarcome cell line usually used as experimental model for *in vitro* biocompatibility studies due to its osteoblastic properties as production of mineralized matrix, high alkaline phosphatase, PTH receptors, and osteonectin presence. MGHA and MGHA-Levo scaffolds were soaked in 15 mL of DMEM at 37ºC and orbital shaking at 200 r.p.m. obtaining the supernatants with the Levo dose released after 3 and 24 h. Controls of DMEM without scaffolds were also prepared. All these media were added to cultured human Saos-2 osteoblasts and cells were incubated for 24 h. Then, cells were washed with PBS, harvested using 0.25% trypsin-EDTA solution, and counted with a Neubauer hemocytometer for the analysis of cell proliferation. Cells were then centrifuged at 310 g for 10 min and resuspended in fresh medium for the analysis of viability, cell cycle, apoptosis, intracellular reactive oxygen species (ROS) content, cell size, and complexity by flow cytometry as described below.

**2.6.1 Flow cytometry studies**

After incubation with the different probes, the conditions for the data acquisition and analysis were established using negative and positive controls with the CellQuest Program of Becton Dickinson. For statistical significance, at least 10,000 cells were analyzed in each sample.

**2.6.2 Cell viability and intracellular reactive oxygen species (ROS) content:** After detachment of Saos-2 osteoblasts, cell suspensions were incubated with 100 μM 2´,7´-dichlorofluorescein diacetate (DCFH/DA) at 37ºC for 30 min. DCF fluorescence was excited at 488 nm and measured with a 530/30 nm band pass filter in a FACScalibur Becton Dickinson flow cytometer. Cell viability was determined by addition of propidium iodide (PI: 0.005% in PBS) to stain the DNA of dead cells. The fluorescence of PI was excited at 488 nm, and the emission was measured with a 670 nm LP in the cytometer described above.

**2.6.3 Cell cycle analysis and apoptosis detection:** Cell suspensions were incubated with Hoescht 33258 (Poly Sciences, Hoechst 5 μg/mL, ethanol 30%, and BSA 1% in PBS), used as a nucleic acid stain, for 30 min at room temperature in darkness. The Hoechst fluorescence was excited at 350 nm, and the emission was measured at 450 nm in a LSR Becton Dickinson flow cytometer. The cell percentage in each cycle phase: $G_0/G_1$, S, and $G_2/M$ was calculated with the CellQuest Program of Becton Dickinson, and the SubG1 fraction (cells with fragmented DNA) was used as indicative of apoptosis.

**2.6.4 Cell size and complexity:** After detachment of Saos-2 osteoblasts, foward angle (FSC) and side angle (SSC) scatters were evaluated as indicative of cell size and complexity respectively using a FACScalibur Becton Dickinson flow cytometer.

**2.7 Statistics**

All data are expressed as means ± standard deviations of a representative of three experiments carried out in triplicate. Statistical analysis was performed using the Statistical Package for the Social Sciences (SPSS) version 19 software. Statistical comparisons were made by analysis of variance (ANOVA). Scheffé test was used for post hoc evaluations of differences among groups. In all of the statistical evaluations, $p < 0.05$ was considered as statistically significant.

**3. RESULTS AND DISCUSSION**

**3.1 Characterization of the MGHA-Levo scaffolds**

A deep morphological and structural characterization of 3D MGHA-Levo scaffolds was carried out with different techniques and is summarized in **Figure 1.** By different SEM microscopies, a high and regular level of hierarchical porosity from macro to mesoporous ranges are shown. In this sense, different scales of porosity can be observed: (i) ultra-large macropores of ca. 450 μm; (ii) macropores with diameters of ca. 80 μm interconnected and (iii) highly ordered mesopores in 2D hexagonal (p6mm, plain group) structure with diameters of ca. 9 nm. These results evidence the hierarchical structure of MGHA-Levo scaffolds. A study by mercury intrusion porosimetry was carried out, confirming also that the impregnation process does not affected to the macroporous properties of these scaffolds (data not shown). To confirm qualitatively the presence of Levo, 3D MGHA-Levo scaffolds were characterized by FTIR (**Figure 1**, bottom). We can observe characteristic bands of Levo molecule (which are indicated by arrows) to 3265 $cm^{-1}$ due to carboxylic group, 2931 $cm^{-1}$ due to alkanes group stretching, 1724 $cm^{-1}$ due to stretching of carbonyl group, 1294 $cm^{-1}$ due to stretching of amines, between 1100 to and 1400 $cm^{-1}$ due to the presence of halogen group. Once confirmed the presence of drug in MGHA-Levo scaffolds, a study by elemental chemical analysis was carried out to calculate the amount of antibiotic loaded by each scaffold. Moreover, the textural properties of these scaffolds were analyzed by $N_2$ adsorption porosimetry with the purpose of test if the drug had been incorporated into their mesoporous structure. Table 1 shows the percentage (%) of Levo in each MGHA-Levo scaffold and the variation of their textural properties after impregnation process. Concerning to the yielding of the loading process, MGHA-Levo scaffolds loaded a 3% of levofloxacin respect to the total drug concentration used in the impregnation solution. Although a priori, this percentage would seem low, it is within the normal range of loading of a drug by the impregnation method [42]. $N_2$ adsorption studies (**Figure 1** and table 1) show a significant decrease in the textural properties, from values of 123 $m^2/g$ and 0.2 $cm^3/g$ for specific surface area and total pore volume, respectively, to 40 $m^2/g$ and 0.1 $cm^3/g$ before and after drug loading. Also, a slight variation in pore diameter was observed, exhibiting a value of 9.3 nm in the MGHA-Levo scaffolds versus 10 nm in the MGHA scaffolds without Levo. It is known that the decrease in the values of specific surface area, total pore volume and pore diameter after drug loading process is related to the confinement of the drug into the mesoporous structure according with the literature [42].

Finally, to confirm at atomic level if the presence of the drug affects to the crystallinity of the apatite phase of MGHA scaffold and if the drug levofloxacin has been loaded into the mesoporous structure in crystalline way, XRD studies have been carried out. **Figure S2** shows the XRD pattern at low and wide scattering angles corresponding to a MGHA-Levo scaffold. Low angle XRD pattern (left) shows a well-defined diffraction maximum at 2θ= 0.86 degree and wide maxima around 2θ= 1.43 and 1.67 degree, which can be indexed as 10, 11 and 20 reflections of a 2D-hexagonal structure with P6mm plane group. Wide angle XRD pattern (right) reveals the presence of nanocrystalline apatite phase exhibiting (002), (211) and (310) reflections. These results highlight that drug incorporation into the mesoporous structure does not affect their structural order, maintaining the 2D-hexagonal structure and the crystallinity of the apatite phase of the pore wall. Moreover, levofloxacin is does not crystallize into mesoporous network, as it has been also shown for other mesoporous matrices and other drugs [43].

### 3.2 *In vitro* levofloxacin release kinetics

The *in vitro* drug release assays from 3D scaffolds were carried out at different pH values in PBS to determine the Levo pH dependent delivery. Thus, pH 5.5 and pH 6.7 were used as representative values of an infectious process, consequence of the bacterial proliferation, and pH 7.4 as value in physiological conditions [23,25]. **Figure 2** displays the Levo release from MGHA 3D scaffolds significantly higher at acid pH values (pH 5.5 and 6.7) than that obtained at physiological conditions (pH 7.4), which follows a slow and sustained drug release. In general, the drug release kinetics from mesoporous matrices are governed, primarily by drug diffusion processes throughout the matrix. Such drug diffusion processes are fitted, generally, to the Higuchi model. However, our results suggest that in addition to the drug diffusion process throughout the mesoporous matrix, a new component is governing the drug release kinetics. Specifically, this new component refers to the MGHA matrix-Levofloxacin interactions at different pH values, as it has been previously reported [44,45]. **Figure 2** also shows a schematic representation of different interactions between silanol (Si-OH) groups of 3D scaffolds and Levo ionizable groups. Levo exhibits different pH-dependent protonation states (cationic, zwitterionic, and anionic forms), given by the 6-carboxylic acid and the N4 piperazinyl group, with pKa values of 6.1 and 8.2, respectively, and with an isoelectric point of 7.1. By other hand, SiOH groups have a pKa value of 4.5. Thus, at pH 5.5, the scaffolds surface exhibits Si-OH groups and deprotonated ones (SiO$^-$) negatively charged. Concerning the levofloxacin molecule, at pH 5.5 is in cationic form because its N4 piperazinyl group (green group) presents positive charge and its carboxylic group (red group) is uncharged. So, in this scenario no interactions between silanol groups and Levo are possible to establish, resulting a fast release of the drug, as it is observed in the release kinetics. This situation changes slightly at pH 6.7, where the most of the Si-OH groups are deprotonated and Levo is in two different protonation states, i.e., in cationic form (similar to the anterior situation) and as zwitterion, coexisting both species. Zwitterionic species have the ability to form hydrogen bonds with silanol matrix, leading to a slower release of the drug at this pH as it has been reported elsewhere [33]. Finally, at pH 7.4, the most Levo molecules are in zwitterionic form, which favors the formation of more hydrogen bonds, showing a slow and sustained release, based on the high stability of the hydrogen bonds compared to other electrostatic interactions. Then, based on these hypotheses, the release profiles can be fitted to a kinetic model, which combines two parameters, i.e. reversible interaction matrix drug as function of pH and drug diffusion process throughout the matrix [46]. Therefore, the theoretical model adopted in this work considers first-order diffusion/convection ($k_s$) and drug association/dissociation ($k_{off}$) following the Eq. 1. This model leads to a decoupling of drug association/dissociation with its diffusion/convection: fast release of initially free drug molecules via diffusion/convection and slow release of bound drug molecules, that is dictated by the dissociation process.

$$\frac{Q_t}{Q_0} = \frac{k_{off}}{k_{on}+k_{off}}\left(1 - e^{-k_s t}\right) + \frac{k_{on}}{k_{on}+k_{off}}\left(1 - e^{-k_{off} t}\right) \qquad \text{Eq. 1}$$

where $Q_t/Q_0$ correspond to accumulative amount of drug released with respect to the amount of drug initially presents in the matrix, per unit time. $k_s$ is the diffusion constant and, $k_{off}$ and $k_{on}$ correspond to dissociation/association constants, respectively. The drug that does not interact with the matrix is governed mainly by $k_s$, whereas that the drug-matrix interaction is governed by $k_{off}$. In this model, the free energy difference ($\Delta G$) between drug free and drug bound to the matrix can also be calculated. This parameter determines the initial burst effect and, it is expressed in the Eq.2.

$$\Delta G = -k_B T \ln\left(\frac{k_{on}}{k_{off}}\right) \qquad \text{Eq. 2}$$

where $k_B$ is the Boltzmann constant and, T is the system absolute Temperature in Kelvin grades. Fitting experimental release patterns to Eq.(1) allowed the determination of the experimental values for $k_s$, $k_{on}$ and $k_{off}$. Then, $\Delta G$ was calculated from Eq.(2). The experimental results are summarized in Table 2. The obtained results show a notable increase in $k_s$, $k_{off}$ and $\Delta G$, by decrease of pH. The $k_{off}$ value, at pH 7.4 is very low, which means that most of the drug is bonded to the matrix through hydrogen bonds, having a small amount of drug dissociated and therefore, low diffusion constant. After a pH decrease (pH 6.7), the drug-matrix unions start to dissociate as consequence of changes in the Levo protonation state. This results in an increase of the $k_s$ and $k_{off}$ constants and of the $\Delta G$ parameter, due to there is a higher proportion of free drug. Finally, the more significant increase, in all the parameters, was at pH 5.5. In this case, no Levo-matrix interactions are possible, obtaining the higher amount of Levo released. In this study, Levo release from MGHA-Levo scaffolds is pH-dependent, since drug-carrier interaction and subsequent release can be modulated by pH changes. Specifically, drug release increases as decreasing pH. Also, the drug-carrier interaction is reversible, allowing Levo to be released in a sustained and/or controlled manner. Nowadays, one of the main risks of the drug sustained release over time is the appearance of antibiotic-multiresistant strains [19,45,46]. Fortunately, the higher amount of Levo released from MGHA scaffolds is produced as the pH decreases, and it only occurs when there is a high bacterial proliferation. Therefore, the pH-dependent Levo release from our systems could minimize the risk of antibiotic resistant strains [46].

**3.3 *In vitro* antimicrobial activity**

The 3D MGHA-Levo scaffolds microbiological effectiveness was determined studying the *S. aureus* growth inhibition. **Figure 3** displays Levo dose daily released at pH 7.4 and pH 5.5 and its microbiological activity against this strain after 7 days of incubation. The MIC in these conditions was of 0.06 μg/mL, which is according to the MIC values for *S. aureus*, obtaining by other authors [45]. The results evidence a Levo daily dose higher than the MIC, during 7 days of incubation. After this time, colony forming unit (CFU) of this bacterial strain grew in the agar plates, confirming loss of antimicrobial activity (data not shown). A local antibiotic administration system provides therapeutic benefits when it ensure a drug concentration adequate in the tissue. This requires that the drug dose daily released is several times higher than the MIC of the microorganism causing the infection [45]. Levo dose (μg/mL) daily released from MGHA-Levo scaffolds is, during 7 days, higher than the MIC for the bacterial strain studied, thus, this system could be effective to prevent an implant-associated infection. A direct assay by confocal microscopy has been carried out for demonstrating the microbiological effectiveness of the MGHA-Levo scaffolds. **Figure 3** (bottom) details comparatively the MGHA scaffolds surface without Levo (left) and MGHA-Levo scaffolds (right) after 7 days of incubation with a *S. aureus* concentration [$10^8$ bacteria/mL] daily renewed. Results show the MGHA scaffolds surface without Levo completely covered of mantle of *S. aureus*. By the contract, the MGHA-Levo scaffolds surface shows total absence of bacteria demonstrating their effectiveness after 7 days of incubation.

**Figure 4** summarizes the obtained results to determine the effectiveness of Levo containing 3D MGHA scaffolds against *S. aureus* biofilm, showing almost the total biofilm destruction after 3 days of incubation. Before treatment (left image), the confocal image shows the biofilm surface formed by lived bacteria (green) covered with mucopolysaccharide coating (blue). After 3 days of treatment, the scenario is significant different showing the almost the total of bacteria dead (red), demonstrating the effectiveness of these 3Dscaffolds against gram–positive biofilm bacteria.

It has been demonstrated that most osteomyelitis are caused by *S. aureus*, however they can coexist with other bacteria type with different etiology as *S. epidermidis* [24]. Moreover, pathogenic characteristics of collection organisms are different to that from clinical isolates (which can have up to 20 % more genes [47]). Therefore, although the results obtained in this study are promising, it is important to remark that it is only a preliminary assay that need further in vitro research (and obviously in vivo one) before considering these results orientated for all important etiologies.

*3.4 In vitro MC3T3-E1 preosteoblast and S. aureus co-culture assays*

To evaluate the multifunctional capability concerning to both antimicrobial activity and bone regeneration in a OM scenario, *in vitro* MC3T3-E1/*S. aureus* co-culture assays were carried out. For this purpose, MC3T3-E1 osteoblast-like cells and *S. aureus* were seeded on both MGHA and MGHA-Levo 3D scaffold surfaces. The concentrations used were of $10^8$ cfu/mL *S. aureus* (suspended in TSB) and $10^4$ MC3T3-E1 osteoblast-like cells/mL (suspended completed medium DMEM). Figure 5 provides two series of images (MGHA and MGHA-Levo 3D scaffolds) of the time progress followed in co-culture assays highlighting the expected effects of the material in each case. In the case of MGHA scaffolds without Levo, at longer times (24h), apoptotic cells appear as rounded red spots while the biofilm maturation keeps in successful progress until a fully developed state and osteoblasts complete disruption and disappearance. In the case of co-cultures on MGHA-Levo scaffold, the outcomes are much more inspiring than before. The confocal images after 6 h show the absolute absence of bacterial population adhered to the material surface and osteoblasts healthily attached and growing. Finally, posterior time images after 24 h and 3 days demonstrate, the scaffold tissue integration in surfaces clean of bacteria, evidencing again a great biocompatibility and adequate cell colonization, as it has also been demonstrated for others 3D systems with coadministration of vancomycin and BMP-2 on *in vitro* co-culture [38]. On the other hand, lactate dehydrogenase (LDH) test was performed to (i) assess the cytotoxicty caused by the biomaterial itself (MGHA and MGHA-Levo 3D scaffolds) and (ii) evaluate the status of preosteoblasts under the co-culture conditions. These results (data not shown) show a significant increased LDH released in MGHA scaffold co-culture conditions, which increases with time. This is due to the coexistence of cells and bacteria, as the later ones have a cytotoxic effect on the osteoblast. Moreover, negligible values of LDH released both in osteoblasts monoculture on MGHA and MGHA-Levo scaffolds are observed, proving no cytotoxic effect of both materials. Finally, results from MGHA-Levo scaffolds indicate a high efficiency of the system for keeping preosteoblasts viability in co-culture conditions with bacteria.

**3.5 *In vitro* biocompatibility studies of levofloxacin dosages**

In an ideal delivery system, biocompatibility represents a basis ground of requirements to be combined with the drug antimicrobial effectiveness and adequate drug release kinetics, originating true multifunctional platforms for tissue regeneration. Biocompatibility studies should be carried out with the cell types with which the system comes into contact, analyzing the effects produced by the material physicochemical characteristics and the concentration of antimicrobial agent released. Recently, MGHA scaffolds have demonstrated excellent properties as highly bioactive behavior, enhanced biocompatibility, no induction of inflammatory response as well as good preosteoblast adhesion, colonization, proliferation and differentiation (supporting information, **Figure S1**). These properties together with the results of Levo release shown in this study, suggest their great potential as multifunctional biomaterials for both bone tissue engineering and local drug delivery systems. Since Levo dose released from MGHA-Levo scaffolds is always higher than MIC for *S. aureus*, is important to evaluate possible cytotoxic effects caused by it on human Saos-2 osteoblasts. This cell type was cultured for 24 h in the presence of medium containing the Levo dose released at pH 7.4, after 3 and 24 h. Specifically, studied doses were of 1.8 μg/mL and 7.7 μg/mL, respectively, which corresponding to the maxim doses of Levo released. Note that, the Levo dose released to the characteristic pH of infection (pH 5.5) is 7.5 μg/mL, similar to the maximum dose assayed in these biocompatibility studies.

**Figure 6** shows the effects of Levo release on osteoblast cell cycle analyzed flow cytometry. The cell cycle phases G0/G1, S and G2/M correspond to repose/cytoplasm growth, DNA replication and mitosis phase, respectively. SubG1, corresponding to DNA fragmentation, was used as indicative of apoptosis. As it can be observed, the Levo doses released after 3 and 24 h (1.8 μg/mL and 7.7 μg/mL, respectively) do not alter any phase of the cell cycle G0/G1, S and G2/M of human Saos-2 osteoblasts. Moreover, SubG1 fraction indicates very low levels of cell death by apoptosis in all the experimental conditions assayed.

In order to know the possible cytotoxic effects on the cell size and internal complexity of Saos-2 osteoblasts, these cell parameters were evaluated by flow cytometry through FSC and 90º SSC light scatters, respectively, and are represented in **Figure 7**. These properties are determined in part by plasma membrane, cytoplasm, mitochondria, pinocytic vesicles and lysosomes. The results highlight that no changes of these parameters were observed in Saos-2 osteoblasts cultured with the Levo doses studied.

**Figure 8** represents cell viability and apoptosis percentages, intracellular reactive oxygen species (ROS) content and cell proliferation values of Saos-2 osteoblasts after culture with DMEM containing the Levo dose released after 24 h, i.e. 7.7 µg/mL. No significant differences (p>0.05) are observed in viability and apoptosis of osteoblasts cultured with this dose, compared to control samples and osteoblasts cultured without Levo (MGHA sample). Therefore, Levo dose released after 24 h does not induce cellular dead. Moreover, the effects of this dose (7.7 µg/mL) on osteoblast proliferation and oxidative stress induced were also analyzed.

The results highlight that although the Levo dosage released after the 24 h is not cytotoxic, as it has been indicated above by the results of cell viability (**Figure 8**), cell cycle phases (**Figure 6**), cell size and internal complexity (**Figure 7**), there is a slight but significant delay ($p < 0.05$) in the osteoblasts proliferation after being cultured with medium containing the Levo dosage released after 24 h (**Figure 8**). However, no significant differences ($p > 0.05$) were found when the ROS content of osteoblasts cultured with the medium containing Levo is compared with the ROS values of control samples and Saos-2 osteoblasts cultured with supernatants ofMGHA scaffolds without Levo (**Figure 8**). Thus, these results confirm that the Levo dosage released after 24 h from MGHA scaffolds does not induce oxidative stress on Saos-2 osteoblasts.

**CONCLUSIONS**

Tridimensional nanocomposite scaffolds with a pH-dependent and locally controlled Levo release have been designed. These 3D nanocomposite scaffolds exhibit a sustained Levo delivery at physiological pH (pH 7.4), which increasing notably when pH decreases to characteristic values of bone infection process (pH 6.7 and pH 5.5). This pH-dependent Levo release is able to inhibit the *S. aureus* growth and to destroy a preformed biofilm, without cytotoxic effects on human osteoblasts. *In vitro* co-culture assays with preosteoblasts and bacteria onto the 3D scaffolds surface demonstrate an adequate cell colonization in entire scaffold surface together with its ability to eliminate bacterial contamination. Although further *in vivo* research is necessary, these findings reveal the possible potential of these 3D nanocomposite scaffolds to be used as therapeutic platforms for treatment and prevention of bone infection.

**ACKNOWLEDGMENTS**

MVR acknowledges funding from the European Research Council (Advanced Grant VERDI; ERC-2015-AdG Proposal No.694160). The author also thanks to Spanish MINECO (MAT2013-43299-R, MAT2015-64831-R, MAT2016-75611-R AEI/FEDER, UE). The authors wish to thank the ICTS Centro Nacional de Microscopia Electrónica (Spain), CAI X-ray Diffraction, CAI NMR, CAI Flow Cytometry and Fluorescence Microscopy of the Universidad Complutense de Madrid (Spain) for the assistance.

**TABLES**

**Table 1.** Levo loaded (%) into the mesoporous structure of MGHA-Levo scaffolds. Textural properties: BET surface area ($S_{BET}$), pore volume ($V_P$) and pore diameter ($D_P$) before and after the impregnation process.

| Samples | % Levo | $S_{BET}$ (m$^2$/g) | $V_p$ (cm$^3$/g) | $D_p$ (nm) |
|---|---|---|---|---|
| **MGHA** | - | 123 | 0.2 | 10.0 |
| **MGHA-Levo** | 3 | 40 | 0.1 | 9.3 |

$S_{BET}$ is the specific surface area determined by the BET method. $V_p$ is the total pore volume calculated using the single point method at $P/P_0 = 0.99$. $D_p$ is the pore diameter calculated from the analysis of the adsorption branch of the isotherm using the BJH method.

**Table 2.** Kinetic parameters estimated in the Eq 1. The parameters $K_{off}$(h$^{-1}$), $K_s$(h$^{-1}$), $\Delta G$(10$^{-21}$J) and $R^2$ correspond to drug-matrix dissociation/association constants, the free energy difference and the correlation coefficient, respectively.

| pH | $K_{off}$ (h$^{-1}$) | $K_s$ (h$^{-1}$) | $\Delta G$(10$^{-21}$ J) | $R^2$ |
|---|---|---|---|---|
| 7.4 | 0.002 | 0.121 | 2.71 | 0.997 |
| 6.7 | 0.011 | 0.155 | 4.19 | 0.996 |
| 5.5 | 0.067 | 0.435 | 19.15 | 0.999 |

**Figures**

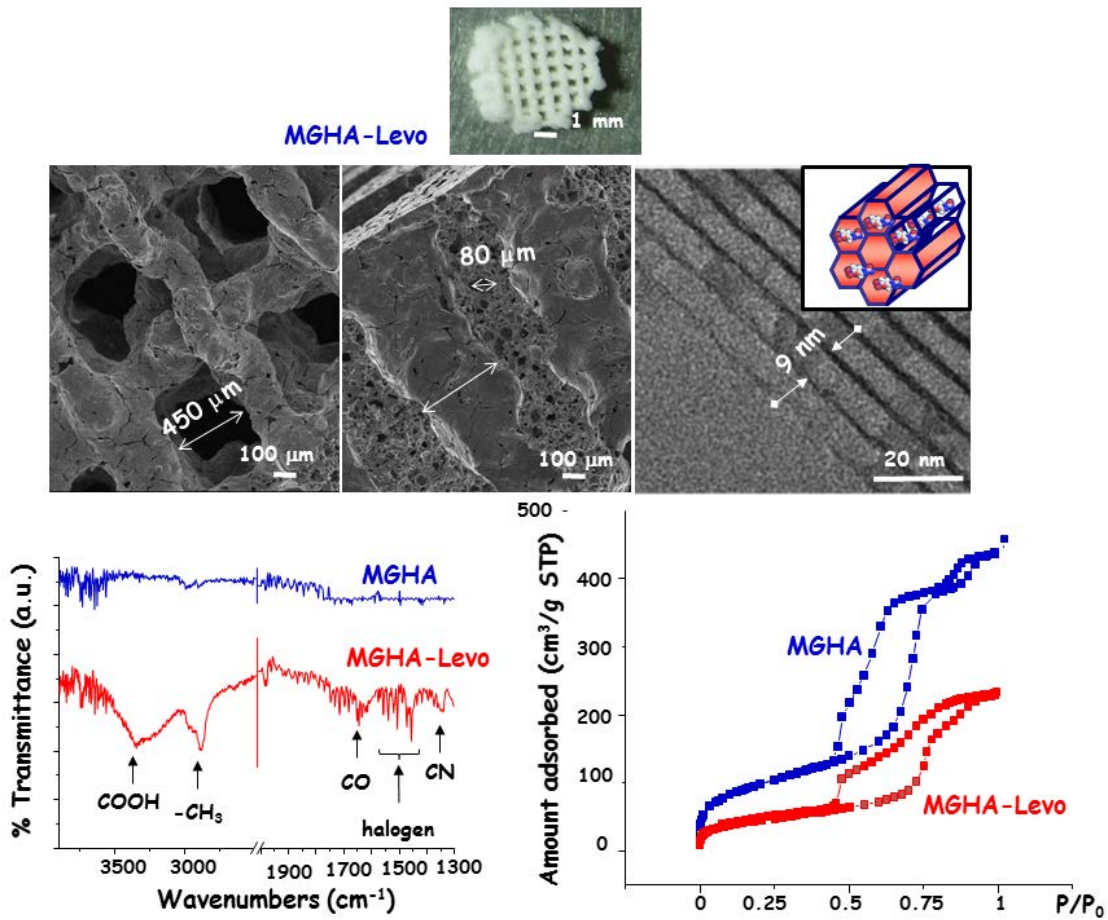

**Figure 1. Characterization of 3D MGHA scaffolds before and after Levo loading.** (Top) Morphological and structural characterization by SEM and TEM showing a high a regular level of hierarchical porosity from macro to mesoporous. (Bottom) FTIR spectra and $N_2$ adsorption isotherms of 3D MGHA scaffolds before and after drug loading confirming the presence of Levo and an decrease of textural properties in the mesoporous range.

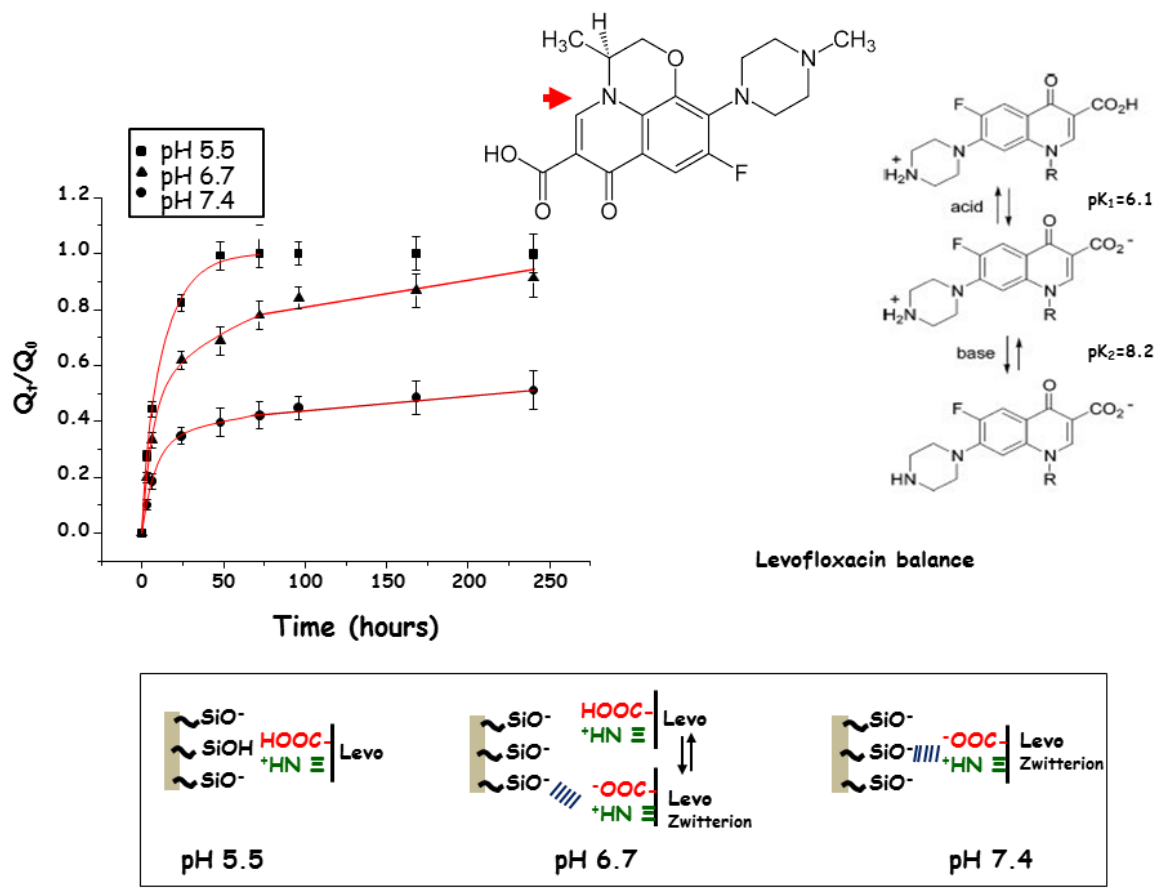

**Figure 2. pH-dependent Levo release from MGHA-Levo scaffolds.** (Left) *In vitro* release profiles to pH 5.5, pH 6.7 and pH 7.4. The curves represent the cumulative amount of Levo released ($Q_t$) with respect to the amount of Levo initially present in the scaffold ($Q_0$) versus time. (Right) Levo acid-basic balance at different pHs. (Bottom) Scheme showing the possible interactions between the silanol groups of the mesoporous matrix and the Levo ionizable groups to different pHs.

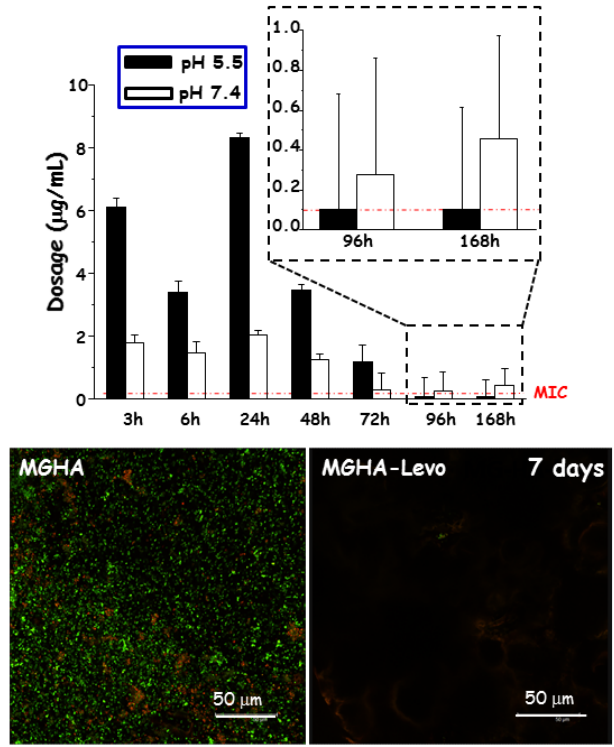

**Figure 3. Levo dose profile daily released.** (Top) Levo dose (μg/mL) released over time to pH 5.5 and pH 7.4. The antimicrobial activity against *S. aureus* has been evaluated in the range of 30 to 0.02 μg/mL, obtaining a value of MIC of 0.06 μg/mL.* [$10^8$ bacteria/mL] daily renewed. (Bottom) A representative image of confocal microscopy.

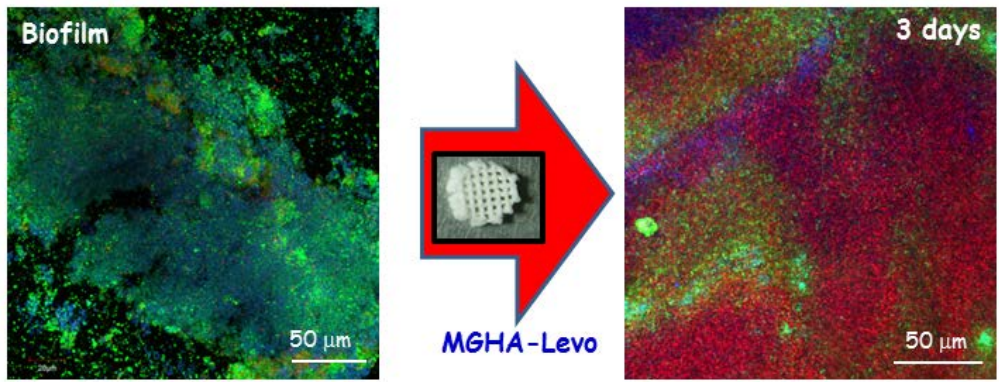

**Figure 4. *In vitro* antimicrobial activity against *S. aureus* biofilm of MGHA-Levo scaffold by confocal microscopy.** (Left) Preformed *S. aureus* biofilm stained by SYTO (Green, alive bacteria) and Calcoflour (blue, mucopolysaccharide covered). (Right) scenario after 3 days with 3D

scaffolds treatment containing Levo showing almost the bacteria stained with IP (read, dead bacteria).

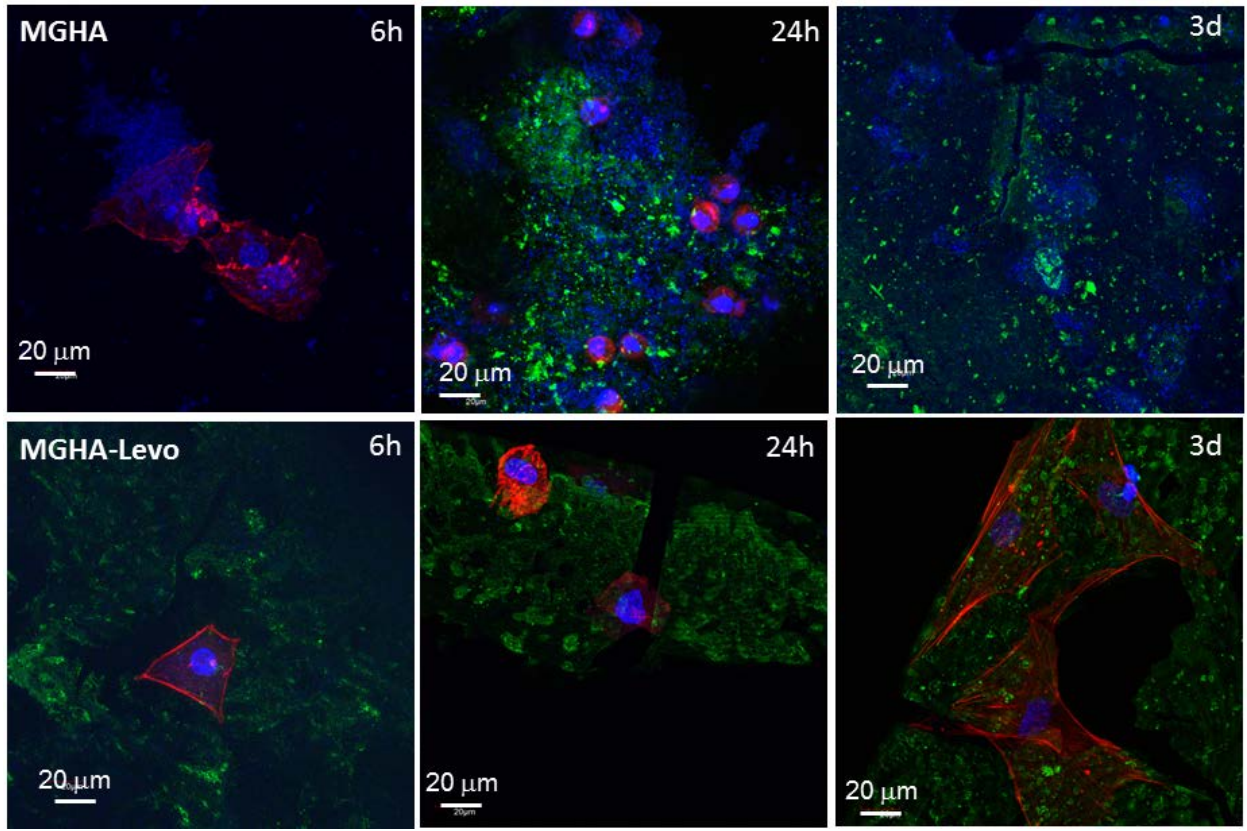

**Figure 5. *In vitro* competitive co-culture MC3T3-E1/*S. aureus* after 6 h, 24 h and 3 days of incubation onto MGHA and MGHA-Levo 3D scaffolds.** Material refraction in green, preostoblastic nuclei and bacteria in blue (DAPI) and actin-fibrous of preosteoblast cytoplasm in red (phalloidin).

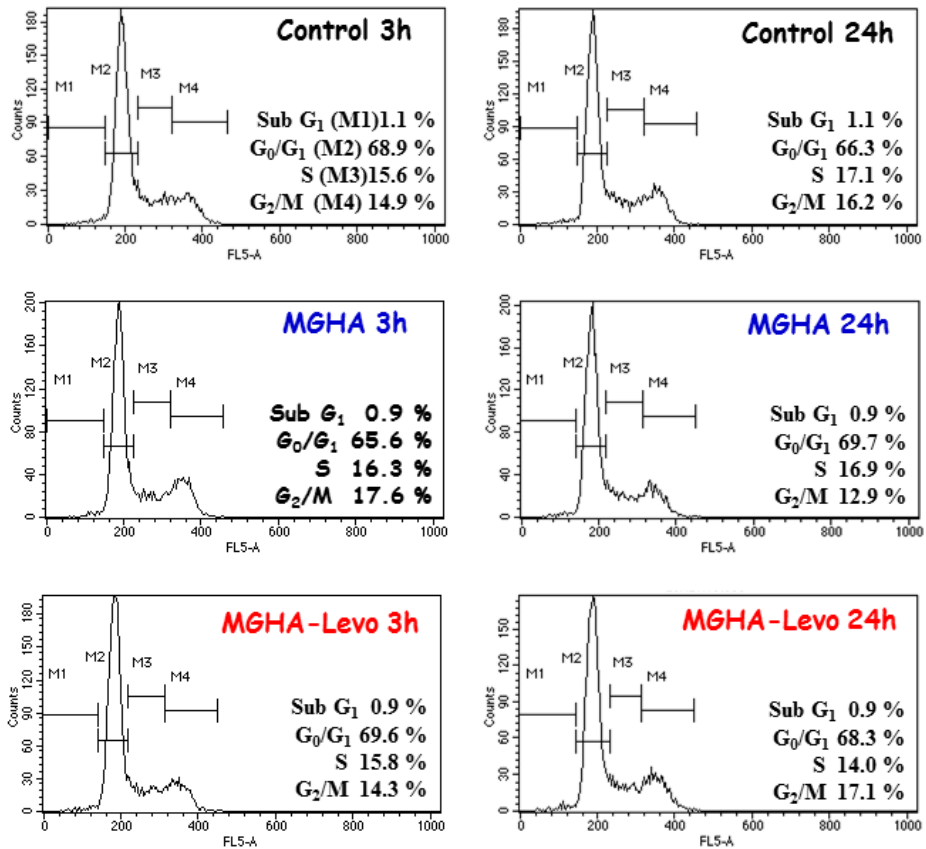

**Figure 6. Effects of Levo dose released from MGHA-Levo scaffolds after 3 and 24 h on the cell cycle of Saos-2 osteoblasts.** Control: osteoblasts cultured in DMEM medium. MGHA 3 h and MGHA 24h: osteoblasts cultured with medium which was in contact with MGHA scaffolds without Levo, after 3 and 24 h, respectively. MGHA-Levo 3h and MGHA-Levo 24h: osteoblasts cultured with

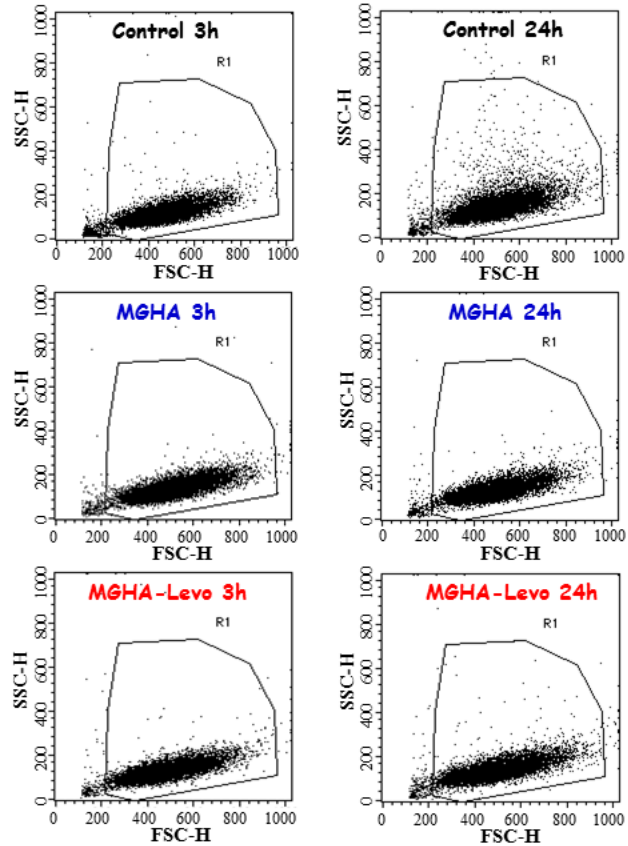

**Figure 7. Effects of Levo dose released from MGHA-Levo scaffolds after 3 and 24 h on the light scattering properties of Saos-2 osteoblasts.** Forward angle scatter (FCS, cell size) versus 90º side angle scatter (SSC, cell internal complexity) after each treatment. Control: osteoblasts cultured in DMEM medium. MGHA 3h and MGHA 24h: osteoblasts cultured with medium which was in contact with MGHA scaffolds without Levo after 3h and 24h, respectively. MGHA-Levo 3h and MGHA-Levo 24h: osteoblasts cultured with medium containing the Levo dose released from MGHA-Levo scaffolds after 3h and 24h, respectively.

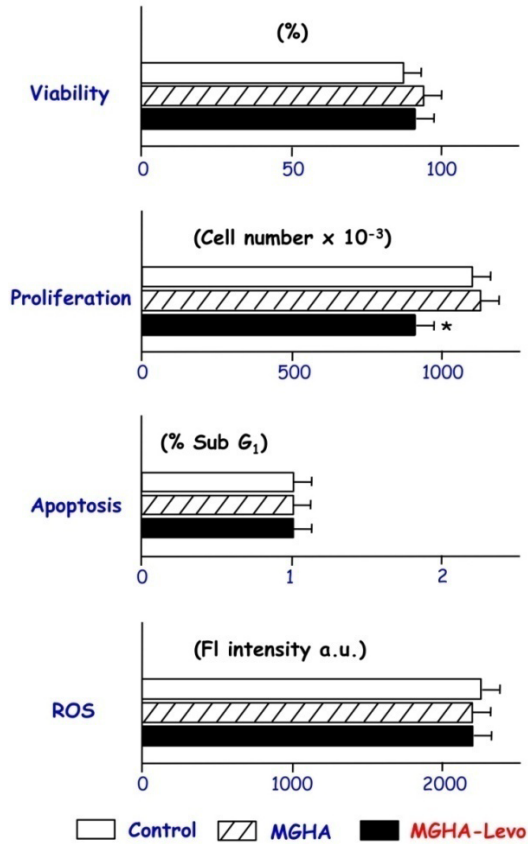

**Figure 8. Effects of Levo dose released from MGHA-Levo scaffolds after 24 hours on viability and apoptosis percentages, intracellular reactive oxygen species (ROS) content and proliferation values of Saos-2 osteoblasts.** Control: osteoblasts cultured in DMEM medium. MGHA: osteoblasts cultured with medium which was in contact with MGHA scaffolds without Levo after 24h. MGHA-Levo: osteoblasts cultured with medium containing the Levo dose released from MGHA-Levo scaffolds after 24h. Statistical significance: *$p < 0.05$.

**Supplementary material**

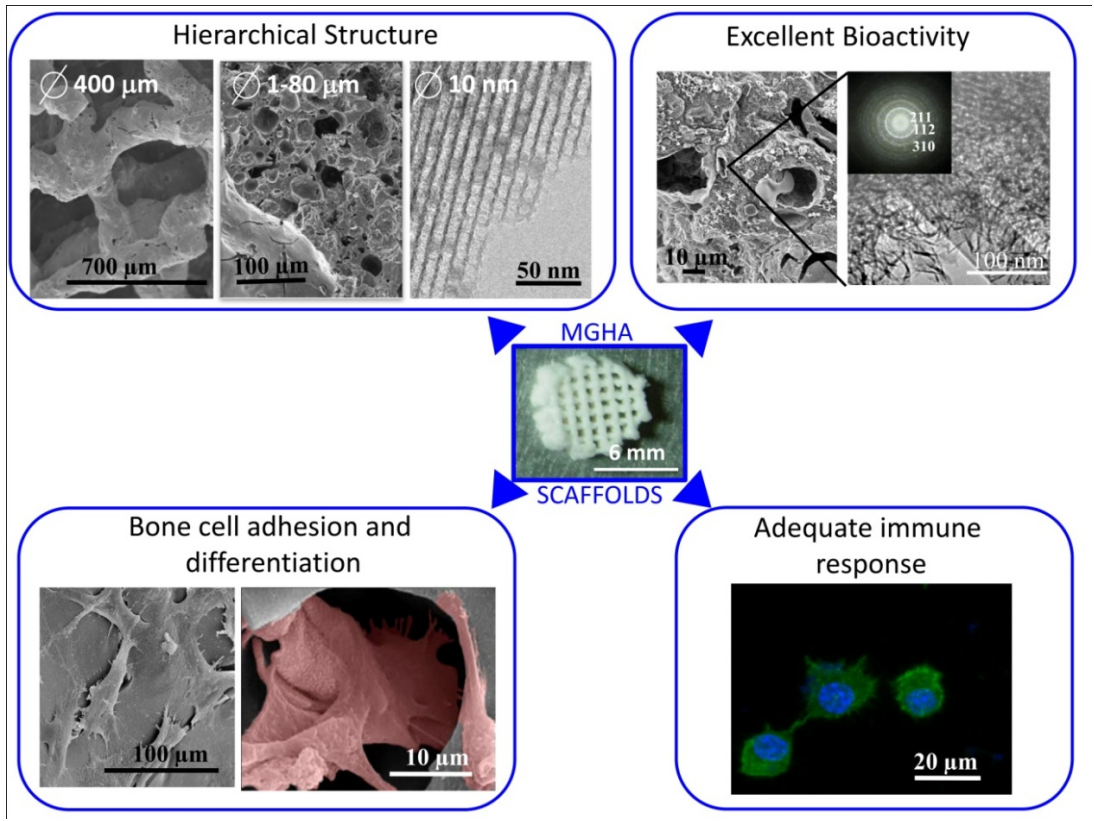

**Figure S1. MGHA scaffolds properties.**

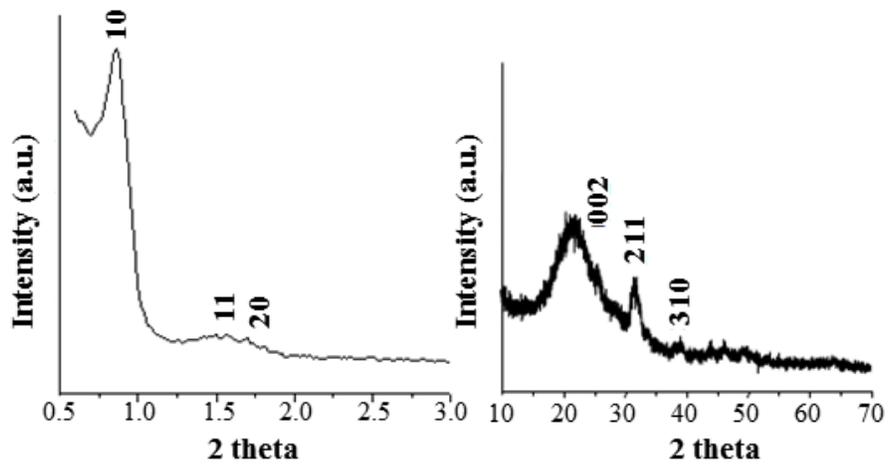

**Figure S2. XRD pattern corresponding to MGHA scaffold at low (left) and wide (right) scattering angles.** The pattern shows, respectively, the ordered mesoporous arrangement and the presence of the apatite phase after Levo impregnation process.

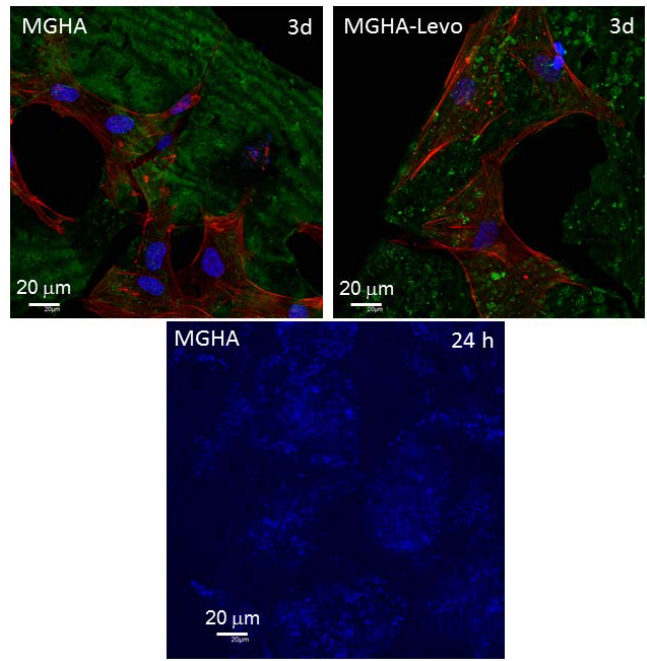

**Figure S3. Confocal microscopy study corresponding to (Top) MC3T3E1 preosteoblastic cells onto the MGHA and MGHA-Levo 3D scaffolds** showing an adequate cell colonization with well-defined anchoring elements on the entire surface of the scaffolds regardless of the presence of antibiotic or not. **(Bottom)** *S. aureus* **seeded onto the surfaces of MGHA (in absence of Levo)** showing the formation of a mantle of bacteria over the entire surface.